\documentclass[final,5p,10pt,a4paper,times,sort&compress]{elsarticle}
\usepackage{ecrc}
\volume{354}
\firstpage{312}
\journalname{Physics Letters A}
\runauth{E. van Groesen et al.}
\jnltitlelogo{\footnotesize PHYSICS LETTERS A}
\renewcommand{\figurename}{Fig.}
\usepackage{latexsym}
\usepackage{graphicx}
\usepackage[fleqn]{amsmath}
\usepackage{amsfonts}
\usepackage{amssymb}
\usepackage{epsfig}
\newtheorem{theorem}{Theorem}

\newtheorem{proposition}[theorem]{Proposition}
\newtheorem{remark}[theorem]{Remark}

\usepackage{etoolbox}
\AtBeginDocument{%
  \patchcmd{\MaketitleBox}{\vspace*{-20pt}\fi}{\fi}{}{}%
}

\usepackage{fancyhdr}
\begin{document}
\begin{frontmatter}
\title{Displaced phase-amplitude variables for waves on finite background}
\author[0,1]{E. van Groesen\corref{cor1}}
\ead{groesen@math.utwente.nl}
\cortext[cor1]{Corresponding author.}
\author[0,1,2]{Andonowati}
\author[0]{N. Karjanto}
\ead{n.karjanto@math.utwente.nl}
\address[0]{Department of Applied Mathematics, University of Twente, The Netherlands}%
\address[1]{LabMath-Indonesia, Bandung, Indonesia}
\address[2]{Department of Mathematics, Institut Teknologi Bandung, Indonesia\\
{\hfill} \\
{\upshape Received 31 January 2006; received in revised form 13 February 2006; accepted 15 February 2006} \vspace{3pt}\\
{\upshape Available online 28 February 2006} \vspace{3pt}\\
{\upshape Communicated by V.M. Agranovich}}

% Abstract
\begin{abstract}
Wave amplification in nonlinear dispersive wave equations may be caused by nonlinear focussing of waves from a certain background. In the model of nonlinear Schr\"odinger equation we will introduce a transformation to displaced phase-amplitude variables with respect to a background of monochromatic waves. The potential energy in the Hamiltonian then depends essentially on the phase. Looking as a special case to phases that are time independent, the oscillator equation for the signal at each position becomes autonomous, with the change of phase with position as only driving force for a spatial evolution towards extreme waves. This is observed to be the governing process of wave amplification in classes of already known solutions of NLS, namely the Akhemediev-, Ma- and Peregrine-solitons. We investigate the case of the soliton on finite background in detail in this Letter as the solution that describes the complete spatial evolution of modulational instability from background to extreme waves. \\
{\footnotesize \copyright \ 2006 Elsevier B.V. All rights reserved.}

\begin{keyword}
Modulational instability \sep Waves on finite background \sep Displaced phase-amplitude \sep Soliton solutions \vspace{5pt}\\

\PACS 46.40.Cd \sep 47.54.Bd \sep 47.35.Bb \sep 47.35.Fg \sep 94.05.Pt \sep 05.45.Yv \sep 52.35.Mw \\
\end{keyword}
\end{abstract}
\end{frontmatter}

% Section 1
\section{Introduction}

Modulational instability is one of the processes that lead to amplification of background waves of small or moderate height into `extreme' waves of large height (rogue or freak waves). The process, different from linear phase focussing, is essentially nonlinear. Since the full nonlinear evolution of the initial linear instability of a monochromatic wave, known as the Benjamin--Feir instability, is difficult to study for the full surface wave equations, we will use the simplified focussing NLS model for a thorough and full investigation.

Despite much recent research~\cite{Osborne00,Henderson,DystheTrulsen,Dysthe,Janssen,Osborne01,Onorato,Andonowati04,vanGroesen04}, it is still rather unclear which circumstances are responsible for the appearance of an extreme wave from a certain background. One of the difficulties seems to be that the actual occurrence of extreme waves cannot be separated from the circumstances that are created by the background. Indeed, following ideas of Boccotti~\cite{Boccotti} it was shown in Craig et al.~\cite{Craig} that the Draupner wave could have developed with linear evolution from the time signal at a much earlier time. In oceanographic situations with forcing by wind and current, a large variety of wave systems that can act as background will be produced. Knowledge of the properties of background wave fields that lead to extreme waves is therefore required. On the other hand, the study of the mathematical--physical properties of extreme waves is equally important since this may help to identify the characterization of backgrounds that are favorable for extreme waves.

In this Letter we address such mathematical--physical aspects for uni-directional waves. Starting with an asymptotic wave field that is a modulated harmonic wave, we will study the full nonlinear evolution that leads to extreme waves. This is done by introducing suitable variables with respect to this background. These will be a kind of displaced phase-amplitude variables, with a displacement that depends on the background. Alternatively, they are the phase-amplitude variables for the complex amplitude of the difference with the background. Most important is that then the dynamics is governed by a Hamiltonian that contains a potential energy from the cubic nonlinearity and from quadratic contributions that depends explicitly on the phase variable itself.

Investigating special cases, we then restrict the phase to depend only on the position and not on time. Then the dynamics at each position is given as the motion of a nonlinear autonomous oscillator in a potential energy that depends on the phase as a parameter and on the spatial phase change. This change of phase with position, which physically corresponds to a change of the wavelength of the carrier wave, turns out to be the only driving force responsible for the nonlinear amplification towards an extreme wave. Remarkably, the assumption that the phase is independent of time leads necessarily to the well-known special solutions of NLS, the soliton on finite background~\cite{Akhmediev85,Akhmediev86,Akhmediev87,Akhmediev}, the Ma-solitons~\cite{Ma} and the algebraic Peregrine solution~\cite{Peregrine}. In this Letter we will restrict to a detailed investigation of the SFB~\cite{Karjanto06}.

This investigation will give us a complete description of the full nonlinear process from small modulations to the extreme wave, and with that of the spatial development of the background in which the extreme wave appears. Calini and Schober~\cite{Calini} investigated this solution (and similar solutions with more unstable sidebands) in a dynamical system approach for persistence as a homoclinic orbit under perturbations. Such results are of paramount importance to study the robustness of the whole generation process.

Our analysis will show that actually a limited number of functionals seem to play a role in the complicated process. Since two of these functionals describe the physically relevant and well defined quantities of (quadratic) energy and Hamiltonian, the optimization properties that we will find may be relevant for other cases too. In fact, these functionals are nowadays also used in statistical descriptions~\cite{Janssen,FedeleArena,Fedele}.

The organization of the Letter is as follows. Section~\ref{2} starts with some preliminaries to fix notation and introduce the variational formulations. Section~\ref{3} defines the transformation to displaced phase-amplitude variables, while Section~\ref{4} specifies for phases independent of time. There the formulation of the evolution of the time signals is given as a family of constrained optimization problems, and the special NLS-solutions are found. The soliton on finite background is described and interpreted in physical variables in detail. The Letter finishes with some remarks and conclusions.

% Section 2
\section{Preliminaries} \label{2}

In this section we introduce the notation to be used, and in particular the variational aspects of NLS that will play a role in the following. We will also show how the most well known solutions of NLS are characterized directly in a variational way.

% Subsection 2.1
\subsection{Variational descriptions for NLS}

We consider real valued wave groups in space ($x$) and time ($t$): $\eta (x,t)$. In lowest order the wave group is described in the standard way as a monochromatic carrier wave with slowly varying complex amplitude $A$ higher order terms that are completely determined by $A$ will be suppressed in the rest of this Letter, so that
\begin{equation*}
\eta(x, t) = A(\xi, \tau)  e^{i\theta_0} + \text{c.c.}, \qquad \theta_0 = k_0 x - \omega_0 t, \qquad k_0 = K(\omega_0),
\end{equation*}
where $\omega \rightarrow K(\omega)$ is the governing linear dispersion relation. To describe the space evolution for the signalling problem for NLS, we use a coordinate system with delayed time: $\xi =x$, $\tau = t - x/V_0$ where $V_0 = 1/K^{\prime}(\omega_0)$ is the group velocity. Then $A$ satisfies the NLS equation which can be written as a Hamiltonian system
\begin{equation*}
i \partial_\xi A + \delta H(A) = 0,
\end{equation*}
where $\delta H$ denotes the variational derivative of the Hamiltonian $H$. The Hamiltonian is given as a functional of functions of $\tau$ by
\begin{equation*}
H = \int \left[\frac{1}{2} \beta |\partial_\tau A|^{2} - \frac{\gamma}{4} |A|^{4} \right]  d \tau.
\end{equation*}
Considering the converging NLS equation, valid for surface waves with sufficiently large carrier frequency $\omega_{0}$, both parameters are positive $\beta >0$, $\gamma > 0$; $\beta$ is related to group velocity dispersion, and $\gamma$ is to the quadratic or third order nonlinearity in the equation for $\eta$. The NLS equation in Hamiltonian form directly result from the action principle
\begin{equation*}
\mathcal{A}(A) = \int \int \frac{1}{2} \left[iA^\ast . \partial_\xi A \right] d\xi d\tau + \int H(A) d\xi
\end{equation*}
Besides the Hamiltonian, the functional $\int|A|^{2} d\tau$ is also conserved; in accordance with~\cite{Akhmediev} this functional will be called the (quadratic) energy. Actually, there are infinitely many invariant integrals, but these two are most relevant for the following and from a physical point of view.

% Subsection 2.2
\subsection{Phase amplitude description}

The description with the complex amplitude $A$ can be translated to phase-amplitude description by introducing polar coordinates:
\begin{equation*}
A = a e^{i\psi}, \qquad a \geq 0.
\end{equation*}
The governing equations are most directly obtained from the action principle, which can be transformed (removing uninteresting total derivatives) into
\begin{equation*}
\mathcal{A}(a, \psi) = \int \int \left[-\frac{1}{2} \left\{a^2 \partial_\xi \psi \right\} + \frac{1}{2} \beta a_\tau^2 + \frac{1}{2} \beta a^2 \psi_\tau^2 - 
\frac{\gamma}{4} a^4 \right]  d\xi d\tau.
\end{equation*}
Variations with respect to $\psi$ leads to the energy equation
\begin{equation*}
\partial_\xi \left[a^2 \right] - \partial_\tau \left[\beta \psi_\tau a^2 \right] = 0
\end{equation*}
while the phase equation results form variations with respect to~$a$:
\begin{equation*}
\beta \partial_\tau^2 a + \left(\psi_\xi - \beta \psi_\tau^2 \right) a + \gamma a^3 = 0.
\end{equation*}
For the following it is relevant to observe that the phase equation is actually an oscillator equation for the amplitude as function of $\tau$:
\begin{equation*}
\beta \partial_\tau^2 a + \left(\psi_\xi - \beta \psi_\tau^2 \right) a + \gamma a^3 = 0.
\end{equation*}
At each position this oscillator equation is non-autonomous in general, and the linear term changes in time and with position through the coefficient $(\psi_\xi - \beta \psi_\tau^2)$, i.e. through the deviation from the linear dispersion relation $K(\omega) - k$.

The conservation of Hamiltonian and quadratic energy now reads
\begin{equation*}
\partial_\xi \int \left[\frac{1}{2} \beta a_\tau^2 + \frac{1}{2}\beta \psi_\tau^2 a^2 - \frac{\gamma}{4} a^4 \right] d\tau = 0, \qquad \qquad
\partial_\xi \int a^2 d\tau = 0.
\end{equation*}

% Subsection 2.3
\subsection{Nonlinear monochromatics and solitons}

As for many PDEs, special solutions can be obtained by separation of variables.

The nonlinear monochromatic solution changes the linear chromatic solution only in the dispersion relation; for a monochromatic wave of amplitude $r_0$ it is the solution
\begin{equation*}
A = r_0 e^{-i \alpha \xi} \quad \text{with} \; \alpha = \gamma r_0^2
\end{equation*}
corresponding to the physical solution (in lowest order) $\eta = 2 r_0 \cos \left((k_0 + \alpha) x - \omega_0 t \right)$. This shows that at the same frequency the wavenumber is modified, according to the nonlinear dispersion relation: $k = K(\omega_0) + \gamma a^2$ with $a = 2r_0$ the amplitude. This monochromatic wave will be used as a  `background' in the next sections. Other solutions of the form
\begin{equation*}
A = S(\tau) e^{-i \alpha \xi}
\end{equation*}
can also be obtained. Substituting this expression in the equation there results as equation for the profile $S$:
\begin{equation*}
\beta S_{\tau \tau} - \alpha S + \gamma S^3 = 0.
\end{equation*}
Solutions exist for each $\alpha > 0$: it is the oscillator equation with effective potential energy $V(S) = -\alpha S^2/2 + \gamma S^4/4$.
Since $\alpha > 0$, the origin is unstable, while boundedness is assured by the fourth order term. The equilibrium is the nonlinear monochromatic above; oscillations around this equilibrium describe modulations. The soliton is the homoclinic solution with $S(\tau) \rightarrow 0$ for $\tau \rightarrow \pm\infty$. The motion in the Argand plane is on a straight line through the origin with angle $-\alpha$ that depends on the position $\xi(= x)$.

For later comparison, it can be observed that the governing equation for $S$ can be found by substituting the Ansatz $A = S(\tau) e^{-i \alpha \xi}$ directly in the action principle, leading to
\begin{equation*}
\int \int \left[\frac{1}{2} \alpha S^2 + \frac{\beta}{2} S_\tau^2 - \frac{\gamma}{4} S^4 \right] d\xi d\tau;
\end{equation*}
critical points with respect to $S$ leads to the governing equation (rendering a formally divergent value to the integral). Stated differently, this can be interpreted as a constrained variational formulation of the quadratic energy at level set of the Hamiltonian:
\begin{equation*}
\max_{S} \left\{\int \frac{1}{2} S^2 d\tau \; \vert \; H(S) = \gamma \right\},
\end{equation*}
where $\alpha$ is the (reciprocal) Lagrange multiplier. We will see a similar formulation in the next sections.

The physical wave fields corresponding to these solutions are given by
\begin{equation*}
\eta(x, t) = 2 S(t - x/V_0) \cos \left((k_0 - \alpha) x - \omega_0 t \right) + \text{hot}.
\end{equation*}
This shows that the carrier wave has fixed wavelength and period, whereas the solutions we will study in Section~\ref{4} will have wavenumber changing with position. For the soliton, the amplitude vanishes at $\xi = -\infty$, increases to a finite value for increasing $\xi$ and vanishes towards $\xi \rightarrow \infty$. The periodic solutions describe modulations of the envelope around the amplitude value $2 r_0$. For each of these solutions, at all fixed positions the time signal $S$ is the same, with a delay at subsequent positions related to the propagation of the envelope. In contrast, the solutions in Section~\ref{4} will have different time signals at different positions.

% Section 3
\section{Description of Waves on Finite Background} \label{3}

The solutions above are never associated with modulation instability: the soliton as homoclinic orbit has exponential growth and decay at infinity, and the modulated oscillation describes a periodic bounded motion. Modulational instability, with the Benjamin--Feir instability of wave trains in surface water waves as prime example, is commonly associated with finite amplitude wave trains that get amplified by self focussing processes due to modulations in the envelope amplitude. We will restrict in the following to this basic setting of a perturbation of a uniform wavetrain, although other `backgrounds' deserve more attention too.

\begin{figure}[htbp]
\begin{center}
\includegraphics[width=0.9\linewidth]{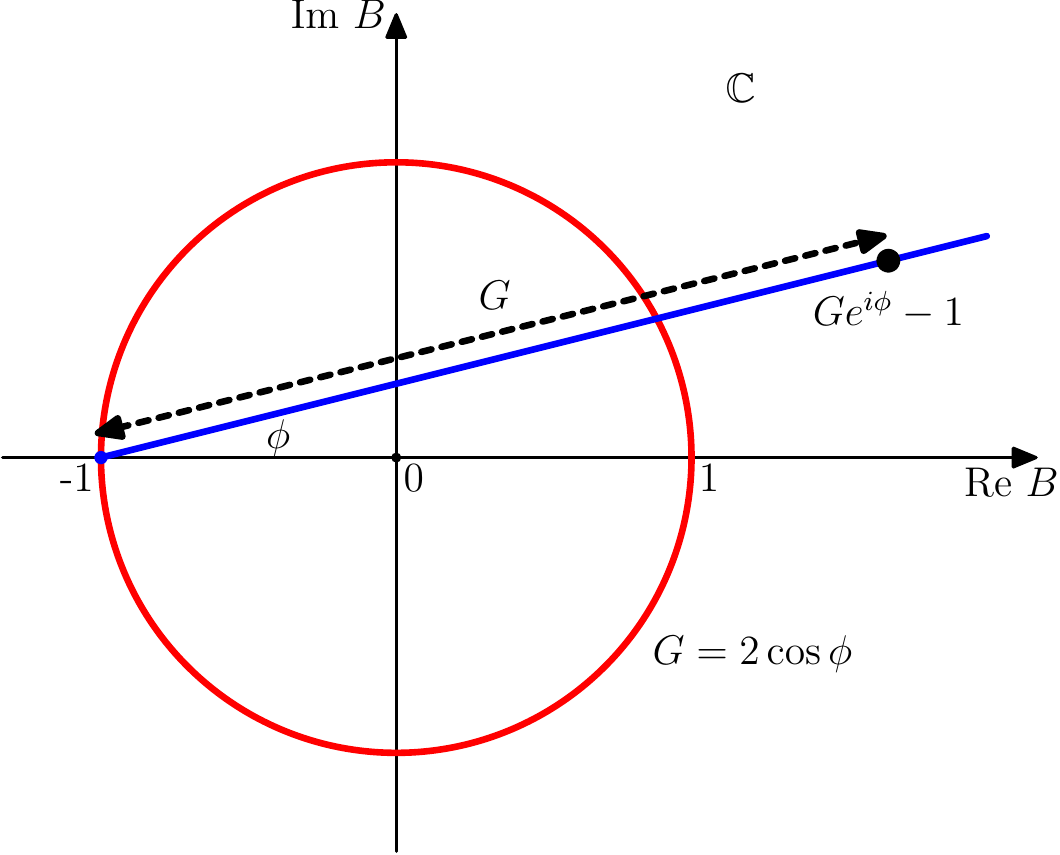}
\caption{\footnotesize In the Argand plane for the NLS amplitude, indicated are the displaced phase-amplitude parameters; the unit sphere corresponds to the set $G = 2 \cos \phi$.} \label{sketsaargand}
\end{center}
\end{figure}

To define more precisely the class of waves we are interested in, we take as background a nonlinear harmonic (plane wave), $r_0 e^{-i \alpha \xi}$, with $\alpha = \gamma r_0^2$, and we will look for solutions of the form
\begin{equation*}
A = B(\xi, \tau) r_0 e^{-i \alpha \xi},
\end{equation*}
where asymptotic properties for $B$ will be specified further. For this asymptotic we will require that except for a possible phase factor, the asymptotic value is the plane wave, i.e. we require that
\begin{equation*}
\left \vert B \right \vert \rightarrow 1 \quad \text{for} \; \xi \rightarrow \pm \infty \; \text{or} \; \tau \rightarrow \pm \infty.
\end{equation*}
This asymptotic behavior motivates, without putting additional restrictions, the introduction of reduced phase $\phi$ and amplitude $G$ parameters according to
\begin{equation*}
B = G(\tau, \xi) e^{i \phi(\xi, \tau)} - 1, \quad G, \phi \; \text{real}.
\end{equation*}
The asymptotic requirement implies that $G - 2 \cos \phi \rightarrow 0$. In the cases below we will deal with solutions for which $G_\tau$ and $\phi_\tau \rightarrow 0$. Then for some limiting phases $\phi_\pm$ it holds that
\begin{equation*}
\phi \rightarrow \phi_{\pm} \quad \text{and} \quad G \rightarrow 2 \cos (\phi_\pm) \; \text{asymptotically}.
\end{equation*}
In the complex (Argand) plane, these parameters are depicted in Fig.~\ref{sketsaargand}.

To arrive at the governing phase-amplitude equations, we substitute $A = \left(G(\tau, \xi) e^{i \phi (\xi, \tau)} - 1 \right) r_0 e^{-i \alpha \xi}$ into the action functional and find:
\begin{equation*}
\mathcal{A}(G, \phi) = r_0^2 \int d \xi \left[\int -\frac{1}{2} \phi_\xi G^2 d \tau + \bar{H}(G, \phi) \right],
\end{equation*}
where the transformed Hamiltonian reads
\begin{equation*}
\bar{H}(G, \phi) = \int \left[\frac{\beta}{2} G_\tau^2 + \frac{\beta}{2} \phi_\tau^2 G^2 - W(G, \phi; \alpha) \right] d\tau
\end{equation*}
and where we have introduced the normalized potential energy as follows. The non-normalized expression for $W$ is given explicitly by $-\alpha |G e^{i \phi} - 1|^2/2 + \alpha \vert G e^{i\phi} - 1 \vert^4/4 = \alpha G^2 (G - 2 \cos\phi)^2/4 - \alpha/4$. In the following we will discard the uninteresting constant term which, moreover, gives rise to divergence of the action functional, and take 
\begin{equation*}
W = \frac{\alpha}{4} G^2 (G - 2 \cos \phi)^2.
\end{equation*}
It is to be noted that $W$ has, besides the origin $G = 0$, the unit circle in the complex plane as set of nontrivial critical points:
\begin{align*}
&\frac{\partial W}{\partial G} = \frac{\partial W}{\partial\phi} = 0 \quad \text{for} \; G = 2 \cos\phi, \\
&\text{for which} \; W = 0 \; \text{and} \; \left\vert Ge^{i\phi} - 1 \right\vert = 1.
\end{align*}

The governing equations follow from variations with respect to $\phi$ and $G$. Variation with respect to $\phi(\xi, \tau)$ gives the (time integrated) energy equation which takes the form
\begin{equation*}
\partial_\xi \left(\frac{1}{2} G^2 \right) - \partial_\tau \left(\beta \phi_{\tau}G^2 \right) - \frac{\partial W}{\partial \phi} = 0
\end{equation*}
while variation with respect to $G(\xi, \tau)$ gives the `displaced' phase equation
\begin{equation*}
\phi_\xi G - \beta \phi_\tau^2 G + \beta G_{\tau \tau} + \frac{\partial W}{\partial G} = 0.
\end{equation*}
The energy equation shows a forcing from the dependence of $W$ on $\phi$. The phase equation is the transformed nonlinear dispersion relation. It can again also be interpreted as a nonlinear oscillator equation for $G$, which now depends on $\phi$ through the dependence in $W$ and on the combination of its derivatives $\phi_\xi - \beta \phi_\tau^2$ that contributes to the coefficient in front of the linear term. In general this is therefore a non-autonomous oscillator equation. Note that it has exactly the same form as the general phase equation, but now $W$ depends essentially on the phase $\phi$. For fixed $\phi$, the energy of each oscillator is therefore not constant in general, but for time periodic motions the time-integral over one period vanishes: there is no nett energy in- or output.

The conservation properties for the spatial evolution are now given by the quadratic energy
\begin{equation*}
\partial_\xi \int \left(G^2 - 2 G \cos \phi + 1 \right) d\tau = 0
\end{equation*}
and the original Hamiltonian%
\begin{equation*}
\partial_\xi \int \left[\frac{\beta}{2} \left(G_\tau^2 + \phi_\tau^2 G^2 \right) - \frac{\alpha}{4} \left \vert G e^{i \phi} - 1 \right\vert^4 \right] d\tau = 0.
\end{equation*}
As a consequence, also the transformed Hamiltonian is conserved: $\partial_\xi \bar{H} = 0$. For solutions with the asymptotic as above, we have $W \rightarrow 0$ for $\xi \rightarrow \pm \infty$. Hence it follows that $\bar{H} = 0$ asymptotically, and then the constancy in $\xi$ implies
\begin{equation*}
\bar{H} = 0 \quad \text{for all} \; \xi.
\end{equation*}

Related to the non-autonomous character of the equation, in general, the dependence of $\phi$ on $\tau$ implies that at a fixed spatial position the motion in the Argand diagram is not on a straight line. In the next section we will consider special solutions for which the motion is at each position represented by a motion on a straight line, the line turning with position.

% Section 4
\section{Pseudo-coherent wave solutions} \label{4}

In this section we consider special solutions for which the displaced phase $\phi$ does not depend on time. In the first section we show that then the spatial evolution is fully described by a family of constrained optimization problems for the time signals. Each optimization problem is only parameterized by the phase, and the spatial dynamics comes in from the change of the multiplier with phase. Remarkably, classes of solutions of these optimization problems can be found explicitly; the corresponding solutions found in this way are then recognized as well-known NLS soliton-type solutions. In the next subsections the specifics are then given for the case of the soliton on finite background, referring to another paper for the other cases.

% Subsection 4.1
\subsection{Parameterized constrained optimization for evolving time signals}

From the assumption that the phase does not depend on time,
\begin{equation*}
B = G(\tau,\ xi) e^{i \phi (\xi)} - 1,
\end{equation*}
it follows that at each position the phase is constant, and the governing oscillator equation for $G$ will be autonomous. Note that although the displaced phase $\phi$ is assumed to depend on $\xi$ only, for the phase of the complex amplitude $B = b e^{i\psi}$, $\psi$ will depend on $\tau$ (and $\xi$), and therefore the NLS signal itself is not coherent; we will call this a pseudo-coherent solution, since it is coherent with respect to the displaced phase-amplitude variables.

The fact that the dependence on $\tau$ is missing in $\phi$ implies that the motions in the Argand plane are on straight lines through the point $-1$, with angle $\phi$ that depends on the position $\xi(= x)$: the solution is displaced over a distance $-1$ in the complex plane.

The equations for $G$ and $\phi$ are a special case of the above, but it is illustrative to derive them directly from the action principle, which now reads
\begin{align*}
&\mathcal{A}(G, \phi) = r_0^2 \int d \xi \left[\int -\frac{1}{2} \phi_\xi G^2 d \tau + \bar{H}(G, \phi) \right] \\
&\text{with} \; \bar{H}(G, \phi) = \int \left[\frac{\beta}{2} G_\tau^2 - W(G, \phi; \alpha) \right]  d\tau
\end{align*}
and with $W$ as above.

\begin{figure}[htbp]
\twocolumn[{%
\begin{center}%
\includegraphics[width=0.4\textwidth]{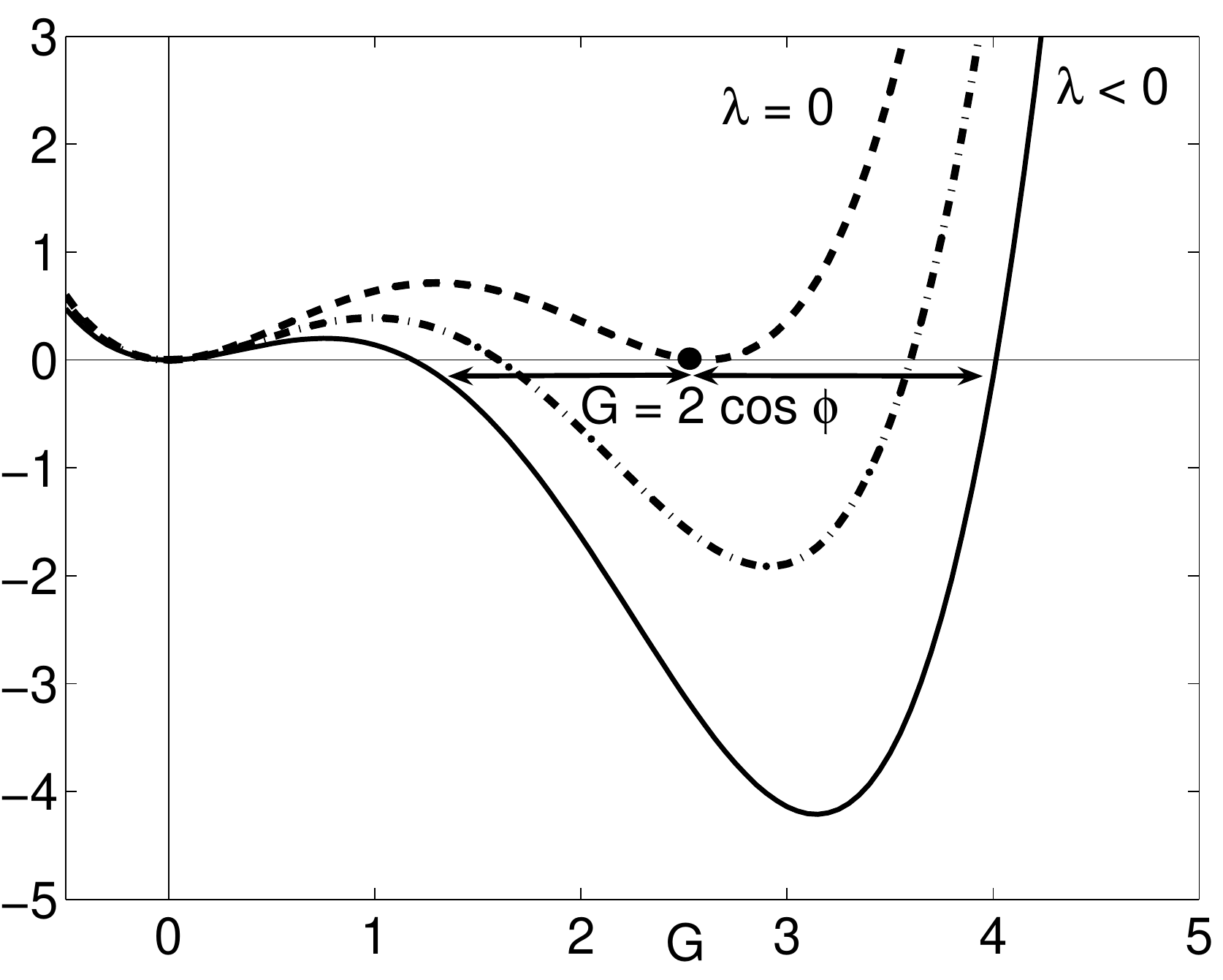} \hspace{1cm}
\includegraphics[width=0.4\textwidth]{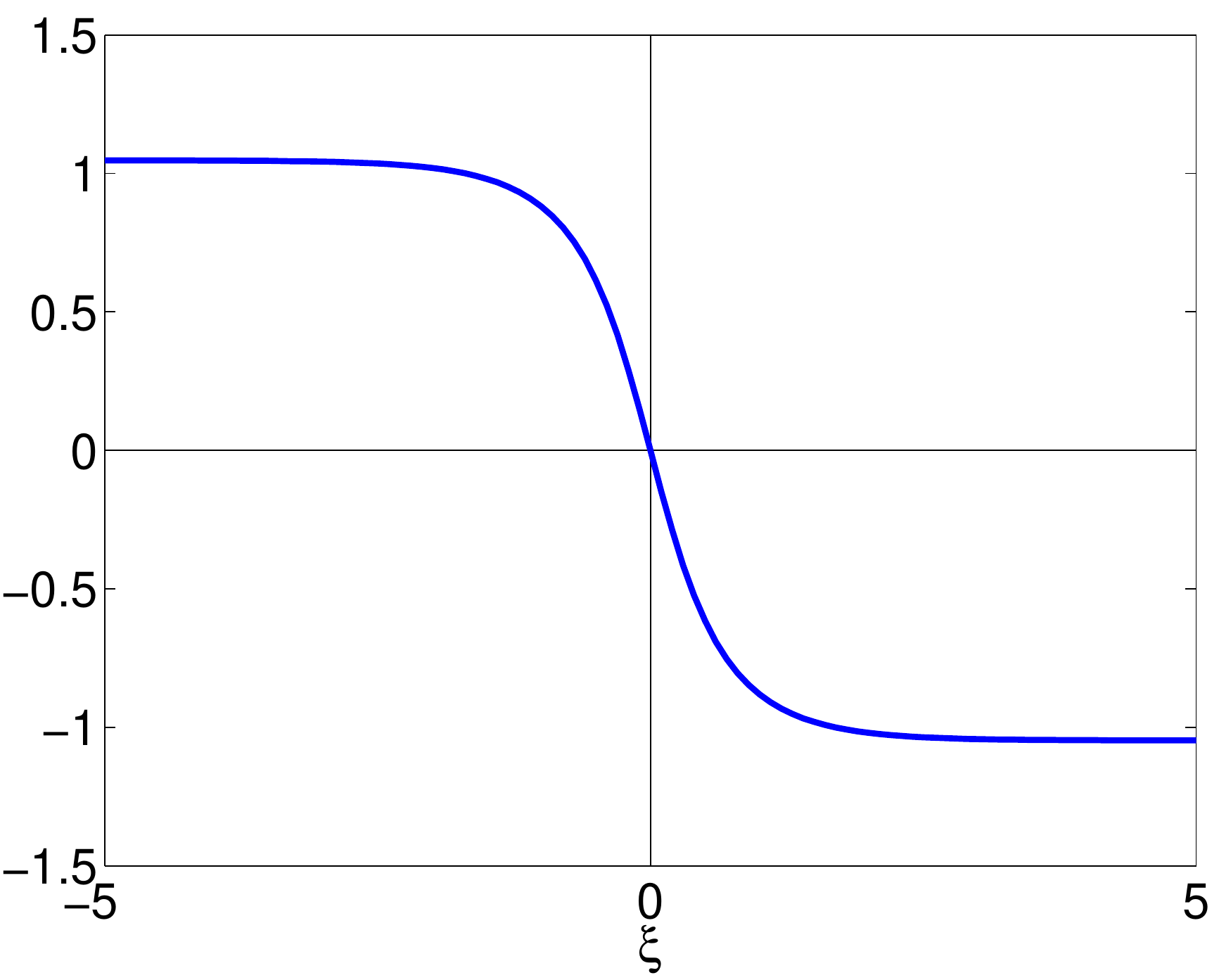}
\end{center}
\caption{\footnotesize (left) Shown are plots of the effective potential for various values of $\lambda < 0$. The dashed line is the plot for $\lambda = 0$. (right) The plot  of the displaced phase of the SFB for $\tilde{\nu} = \sqrt{1/2}$.} \label{potentialphase} \vspace{10pt}
}]  
\end{figure}

Variation with respect to $\phi(\xi)$ gives the time integrated energy equation
\begin{equation*}
\partial_\xi \int \frac{1}{2}(G^2) d \tau - \int \frac{\partial W}{\partial \phi} d\tau = 0
\end{equation*}
which shows the forcing from the dependence of the potential energy on the phase. Variation with respect to $G(\xi, \tau)$ gives the displaced phase equation
\begin{equation*}
\phi_\xi G + \beta G_{\tau\tau} + \frac{\partial W}{\partial G} = 0.
\end{equation*}
This last equation shows that at fixed $\xi$ the nonlinear oscillator equation for $G$ depends on $\phi$ and $\phi_\xi$ but not on $\tau$, so that at each position the equation for $G$ is autonomous. An effective potential energy defined by
\begin{equation*}
V = W + \frac{1}{2} \lambda G^2 = G^2 \left[\alpha^2 (G - 2 \cos \phi)^2/4 + \lambda/2 \right] 
\end{equation*}
is the oscillator potential for $\lambda = \partial_\xi \phi$. For negative $\lambda$ the plots of the potential are depicted in Fig.~\ref{potentialphase}.

The conservation properties for the spatial evolution are as before the quadratic energy and the Hamiltonian; in particular
\begin{equation*}
\bar{H} = \int \left(\frac{\beta}{2} G_\tau^2 - W \right) = 0 \quad \text{for all} \; \xi.
\end{equation*}

Inspection of the oscillator equation shows that the solutions of the oscillator equations can be obtained in the following, somewhat surprising, variational way, the idea of which is motivated by the terms in the action functional.

% Proposition 1
\begin{proposition}
Consider for each $\phi$ the constrained variational problem
\begin{equation}
\text{\upshape Crit}_{G} \left\{\int \frac{1}{2} G^2 \; \vert \; \bar{H}(G, \phi) = 0 \right\}.
\end{equation}
The non-trivial solutions $\tau \rightarrow \bar{G}(\phi)$ satisfy the Lagrange multiplier equation for some (reciprocal) multiplier $\lambda(\phi)$
\begin{equation}
\lambda G = \delta_{G} \bar{H}(G, \phi).
\end{equation}
Then, if $\xi \rightarrow \phi(\xi)$ is a solution of the equation
\begin{equation}
\partial_\xi \phi = \lambda(\phi),
\end{equation}
the spatial evolution $\xi \rightarrow \bar{G}(\phi(\xi))$ leads to a solution of the NLS equation given by
\begin{equation}
A = \left(G e^{i \phi(\xi)} - 1 \right) r_0 e^{-i \alpha \xi}.    \label{WFB_special_ansatz}
\end{equation}
\end{proposition}

The multiplier equation reads in detail
\begin{equation}
\partial_\tau^2 G + \lambda G + \alpha G(G - \cos \phi)(G - 2 \cos \phi ) = 0    \label{fullosceqn}
\end{equation}
and has linear, quadratic and cubic nonlinear terms. It is interesting that various classes of solutions of the constrained variational problem can actually be given explicitly and simply. All solutions are of the form
\begin{equation}
G = \frac{p}{q - \zeta(\tau)}   \label{generalspecialWaveform}
\end{equation}
where $p$ and $q$ depend on $\phi$ and $\zeta(\tau)$ is one of three special functions, each of which corresponds to a well-known special solution. For $\zeta(\tau) = \tau^2$ the solution is the well-known Peregrine (algebraic) soliton~\cite{Peregrine}, while for $\zeta(\tau) = \cosh(\nu \tau)$ the solutions represent the class of Ma-solitons~\cite{Ma}. For $\zeta(\tau) = \cos(\nu \tau)$ the other well-known `soliton on finite background'~\cite{Akhmediev85,Akhmediev86,Akhmediev87,Akhmediev} is obtained. We will present here the results for this last case since this describes the fully nonlinear spatial evolution of the Benjamin-Feir instability of a modulated monochromatic time signal. For the other cases we refer to~\cite{Karjanto06}.

% Remark 2
\begin{remark}
{\upshape In the variational formulation above, it is possible to take another target functional to be optimized:
\begin{equation}
\textmd{Crit}_{G} \left\{\int G \; \vert \; \bar{H}(G,\phi) = 0 \right\}.
\end{equation}
This is a consequence of the fact that the quadratic energy is conserved, implying that
\begin{equation*}
\int \left(G^2 - 2 \cos(\phi) G \right) = \text{constant}.
\end{equation*}
Of course the optimal solutions (and the multiplier) will be different, actually just a shift in $\bar{G}$. This last formulation, for the signal at the extreme position only, was proposed in~\cite{Andonowati04,vanGroesen04}.}
\end{remark}

% Subsection 4.2
\subsection{Soliton on finite background}

This class of solutions is, except for time and space shifts, a one parameter family depending on $\nu$ which is the modulation frequency of the background wave. The details are as follows.

% Proposition 3
\begin{proposition}
For any value of a modulation frequency $\nu$ such that $\tilde{\nu} := \nu \sqrt{\beta/\alpha} < \sqrt{2}$ the solution is given by
\begin{equation*}
\bar{G}(\phi) = \frac{P(\phi)}{Q(\phi) - \cos (\nu \tau)}
\end{equation*}
where the coefficients are given by
\begin{equation*}
P = \frac{\tilde{\nu}^{2} Q}{\cos \phi}, \qquad Q^2 = \frac{2 \cos^2 \phi}{2\cos^2 \phi - \tilde{\nu}^2}
\end{equation*}
and the multiplier is given by
\begin{equation*}
\lambda = (\tilde{\nu}^2 - 2 \cos^2 \phi )/\alpha.
\end{equation*}
Note that all solutions are periodic with the same period $T = 2\pi/\nu$.
\end{proposition}
\begin{figure}[htbp]
\begin{center}
\includegraphics[width=0.8\linewidth]{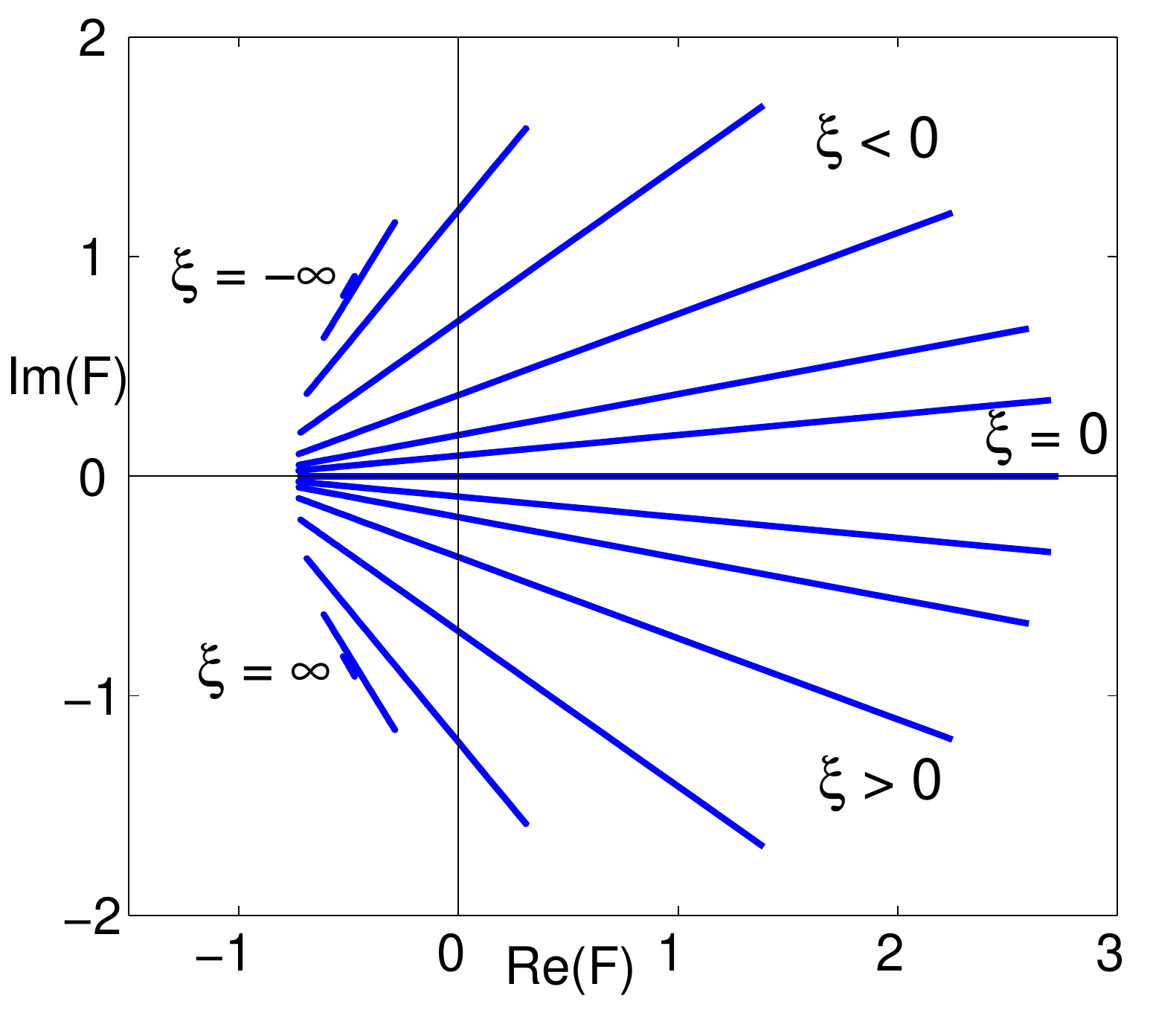}
\end{center}
\caption{\footnotesize The dynamic evolution of the SFB parameterized in $\tau$ at different $\xi$ for $\tilde{\nu} = \sqrt{1/2}$.} \label{ArgandSFB}
\end{figure}

The result can easily be verified by considering the function~\eqref{generalspecialWaveform} with $\zeta(\tau) = \cos(\nu \tau)$:
\begin{equation*}
f = \frac{p}{q - \cos(\nu \tau )}
\end{equation*}
Direct differentiation and algebraic manipulations show that this function satisfies an equation with linear, quadratic and cubic terms. Adjusting $p$, $q$ in such a way that the coefficients are similar to those of the governing equation~\eqref{fullosceqn} leads to the expressions given above. The explicit spatial evolution is then found by solving $\partial_\xi \phi = \lambda = (\tilde{\nu}^2 - 2 \cos^2 \phi)/\alpha$ for $\phi$ as function of $\xi$. The result can be written in elementary functions:
\begin{equation*}
\phi(\xi) = -\arctan \left[\frac{\sqrt{2 - \tilde{\nu}^2}}{\tilde{\nu}} \tanh(\sigma \xi) \right], \quad \text{where} \; \sigma = \alpha \tilde{\nu} \sqrt{2 - \tilde{\nu}^2}.
\end{equation*}
Note that the value of $v$ determines the asymptotic values of the phase:
\begin{equation*}
\tilde{\nu} = \sqrt{2} \cos\phi_{\pm}
\end{equation*}
with $\phi_{+} > \phi_{-}$ to assure that $\phi$ is decreasing. A characteristic plot of this function is given in Fig.~\ref{potentialphase}.

Upon substituting all these results in the expression~\eqref{WFB_special_ansatz} leads after some algebraic manipulations to the solution that is known as the soliton on finite background (SFB), described by~\cite{Akhmediev85, Akhmediev87, Akhmediev}. We will present several observations about the physical process it describes, and the interpretation of the results in the next subsection.

% Remark 4
\begin{remark}
{\upshape The SFB solution found here is modulated by one `sideband' as we shall see. It is the first in an infinite series of solutions which are characterized by the number of sidebands~\cite{Akhmediev}. These SFB solutions are also described in~\cite{Akhmediev85,Akhmediev,Calini,Akhmediev91}.}
\end{remark}

% Subsection 4.3
\subsection{Interpretations}

We will now illustrate the spatial evolution of the modulational instability described by SFB in various ways. It should be kept in mind that only first order effects are shown; when considering the NLS-model for surface wave phenomena, adding the second order bound waves will lead to some quantitative changes.

In the Argand diagram the time evolution is on a straight line through the `background', under an angle that decreases for increasing position; this is illustrated for the case $\tilde{\nu} = \sqrt{1/2}$ in Fig.~\ref{ArgandSFB}. Observe that only for $\xi = 0$, when the trajectory is on the real axis, the time signal is coherent. The special relation between the real and imaginary part for certain solutions of the NLS equation has been described by~\cite{Akhmediev86}.

The next plot, for the same values of the parameters, gives a snapshot of the spatial evolution which we will describe in some detail.
\begin{figure}[htbp]
\begin{center}
\includegraphics[width=0.95\linewidth]{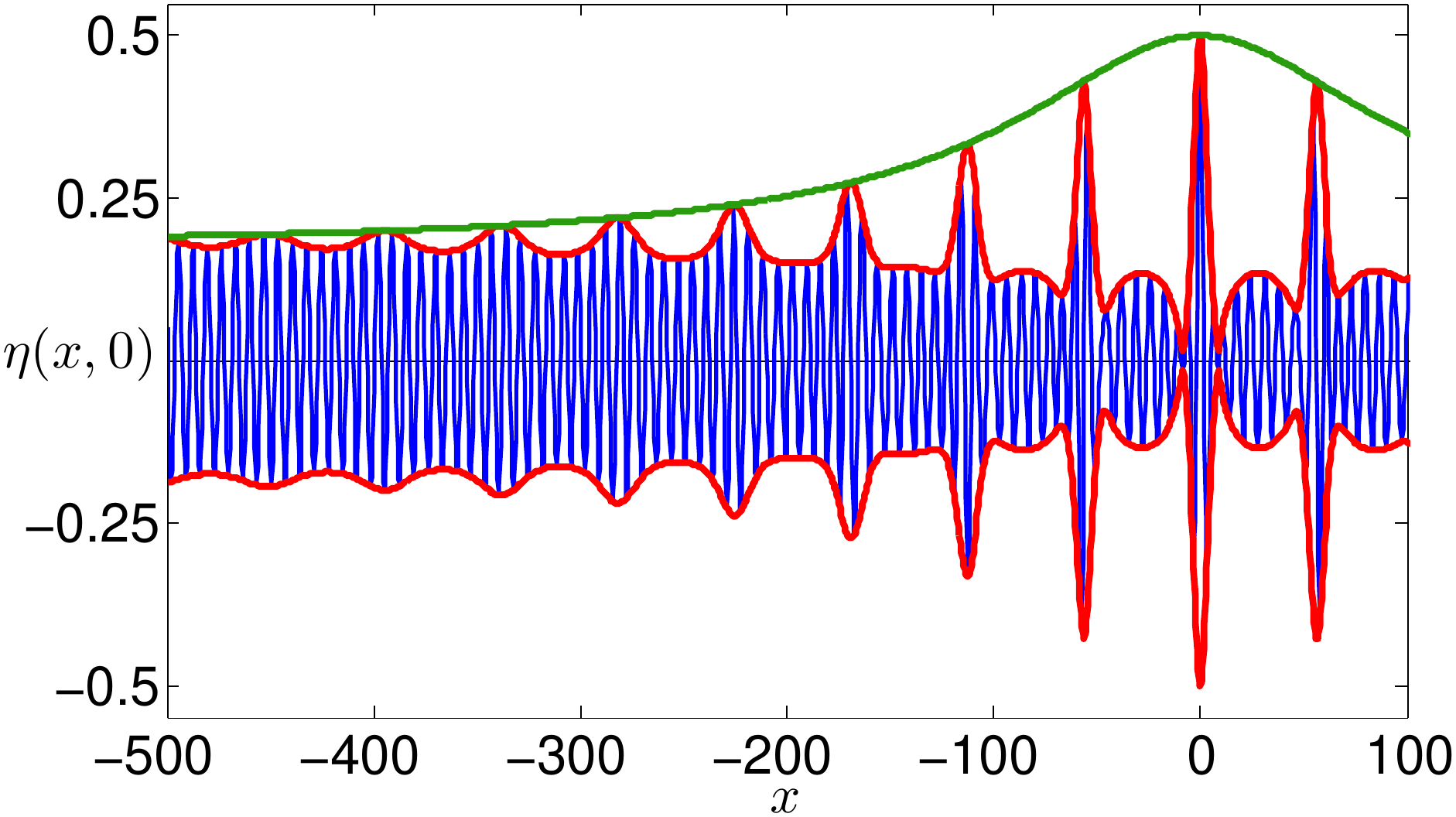}
\end{center}
\caption{Snapshot of the spatial wave field SFB for $\tilde{\nu} = \sqrt{1/2}$ and $\omega_{0} = 3$~rad/sec. Shown are the individual waves, their envelope at that instant, and the time-independent MTA, maximal temporal amplitude. Physical dimensions are given along the axis: horizontally the distance and vertically the surface elevation in meters, for waves on a layer with depth of 5~meter. It is to be noted that this plot only shows the first order waves; adding second order bound waves will deform the waves and increase the MTA.} \label{MTA_SFB}
\end{figure}

Fig.~\ref{MTA_SFB} shows the spatial wave field at some instant of waves, say, running from left to right. At the left the slightly modulated uniform wave train (amplitude $2r_0$) is seen. This modulation clearly determines a characteristic modulation length that is maintained during the complete downstream evolution. While moving to the right, the modulations are amplified, creating distinct wave groups. This initial phase is the so-called Benjamin--Feir instability~\cite{Benjamin}; the exponential growth rate from linear analysis depends on the value of $\nu$ and is given by the expression $\sigma = \sigma(\nu)$ above. After this initial exponential growth of the modulations, nonlinear effects come into play, enforce and later bound the increase in modulation amplitude. Till, at a certain position, called the \textit{extreme position} (in scaled variables taken to be at $x = 0$), the largest wave appears, after which the reverse process sets in the decay towards the asymptotic harmonic wave train (with some phase change). Defining the amplification factor of the whole process as the quotient $Q$ of the highest crest and the background amplitude, the amplification is larger for smaller $\nu$, maximal~$3$ (obtained in the limit $\tilde{\nu} \rightarrow 0$), as follows from the explicit expression given explicitly by~\cite{Osborne00, Karjanto02} $Q = 1 + 2 \sqrt{1 - \tilde{\nu}^2/2}$. Note that the local amplification factor near the extreme position can actually be much larger, since near the extreme position, the extreme wave is locally surrounded by waves of much smaller amplitude, as if the total energy in one wavegroup is conserved but with the energy redistributed between waves.

Dynamically in time, both the waves and the envelope shifts to the right at different speed (the phase and group velocity, respectively). Also shown in the plot is the so-called MTA, the \textit{maximal temporal amplitude}~\cite{Andonowati03}. This is the (time-independent) curve determined by the maximal wave height at each point; it is the steady envelope of the wavegroups. This MTA is easily found from the explicit formulas above, by looking at bounds for $G$ at each position:
\begin{equation*}
2 \cos(\phi) - \frac{\sqrt{-2\lambda(\phi)}}{\alpha} \leq G(\tau, \phi) \leq 2 \cos(\phi) + \frac{\sqrt{-2 \lambda(\phi)}}{\alpha}
\end{equation*}
The MTA then follows from MTA$(\xi)^2 = \max_{\tau} \vert G e^{i \phi} - 1 \vert^2$. Simplification of this expression leads to the result as in~\cite{Andonowati04}
\begin{equation*}
\text{MTA}(x)^{2} = (2 r_0)^2 \left[1 + \frac{2 \tilde{\nu}^2 \sqrt{1 - \tilde{\nu}^2/2}}{\cosh(\sigma x) - \sqrt{1 - \tilde{\nu}^2/2}} \right].
\end{equation*}

\begin{figure}[htbp]
\begin{center}
\includegraphics[width=0.95\linewidth]{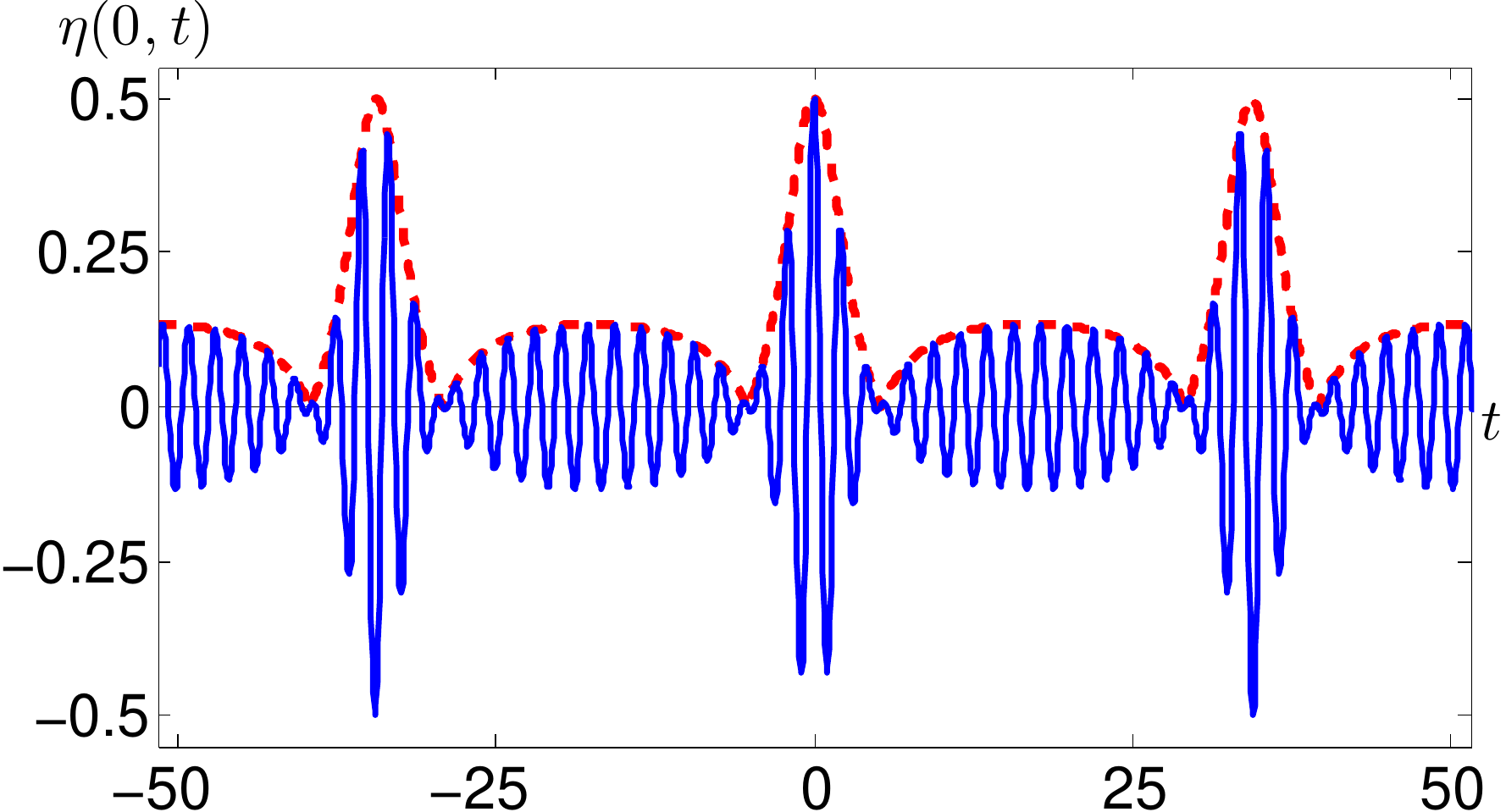}  
\end{center}
\caption{\footnotesize Plot of the SFB signal at the extreme position for $\tilde{\nu} = \sqrt{1/2}$ and $\omega_{0} = 3$ rad/sec.}  \label{Extremesignal}
\end{figure}

In Fig.~\ref{Extremesignal}, the time signal at the extreme position is shown for the same values as above, to illustrate the appearance of phase singularities~\cite{Nye}. These correspond to wave dislocations of the spatial-temporal wave field, wave splitting and merging, and happens when the complex amplitude vanishes. This is the case (only at the extreme position) for sufficiently small values of~$\tilde{\nu}$, explicitly $0 < \tilde{\nu} < \sqrt{3/2}$; see~\cite{Andonowati04,Karjanto07}.

% Section 5
\section{Conclusion and remarks}

We have investigated in this Letter a specific, simple, `finite background' and introduced displaced phase-amplitude variables to look for waves deviating from this background. We found the special known NLS-solutions by assuming the phase not to depend on time explicitly. This leads to a variational characterization of the time signals at each position; the change in phase with position drives the modulational instability process in these cases. The time signal at the position where the extreme waves appear, corresponds to a coherent solution, at other positions this is not the case. Conservation of the quadratic energy and of the Hamiltonian are essential in the described processes. Therefore it may be expected that some of the results can be carried over to models that approximate the full surface wave equations more precisely than NLS to show similar properties. An investigation guided by the formulation with an action principle seems most promising.

As a side result we found the characteristic wave forms for the time signals~\eqref{generalspecialWaveform}; it turns out that these, fully nonlinear profiles are a nonlinear modification of cornered waveforms that are obtained by changing the Hamiltonian to a quadratic expression and looking for waves with maximal crest height; see~\cite{vanGroesen06} for further details for finite energy and~\cite{vanGroesen07} for time periodic signals.

\section*{Acknowledgement} 
This work is executed at University of Twente, The Netherlands as part of the project \textit{``Prediction and generation of deterministic extreme waves in hydrodynamic laboratories"} (TWI. 5374) of the Netherlands Organization of Scientific Research NWO, subdivision Applied Sciences STW.

\end{document}